\documentclass[11pt,a4paper]{article}
\usepackage[hyperref]{acl2021}
\usepackage{amsmath}
\usepackage{times}
\usepackage{latexsym}
\usepackage{enumitem}
\usepackage{alphabeta}
\usepackage{comment}
\usepackage[font=small,skip=1pt]{caption}

\usepackage{subfig}
\usepackage{appendix}
\usepackage{blindtext}

\usepackage{graphicx}

\usepackage{microtype}

\aclfinalcopy 
\def\aclpaperid{3147} 

\title{Cross-replication Reliability - \\An Empirical Approach to Interpreting Inter-rater Reliability\\\normalsize{ACL 2021 Main Conference}}

\author{Ka Wong \\
  Google Research \\
  \texttt{danicky@gmail.com} \\\And
  Praveen Paritosh \\
  Google Research\\
  \texttt{pkp@google.com} \\\And
  Lora Aroyo \\
  Google Research\\
  \texttt{l.m.aroyo@gmail.com}}

\date{06/2021}

\begin{document}
\maketitle
\begin{abstract}
When collecting annotations and labeled data from humans, a standard practice is to use inter-rater reliability (IRR) as a measure of data goodness \cite{Hallgren:2012}. Metrics such as Krippendorff’s alpha or Cohen’s kappa are typically required to be above a threshold of 0.6 \cite{Landis:1977}. These absolute thresholds are unreasonable for \textit{crowdsourced data} from annotators with high cultural and training variances, especially on \textit{subjective topics}. We present a new alternative to interpreting IRR that is more empirical and contextualized. It is based upon benchmarking IRR against baseline measures in a replication, one of which is a novel \textit{cross-replication reliability} (xRR) measure based on Cohen's (1960) kappa. We call this approach the xRR framework. We opensource a replication dataset of 4 million human judgements of facial expressions and analyze it with the proposed framework. We argue this framework can be used to measure the quality of crowdsourced datasets.
\end{abstract}

\section{Introduction}
\label{sec:intro}
Much content analysis and linguistics research is based on data generated by human beings (henceforth, annotators or raters) asked to make some kind of judgment. These judgments involve systematic interpretation of textual, visual, or audible matter (e.g. newspaper articles, television programs, advertisements, public speeches, and other multimodal data). When relying on human observers, researchers must worry about the quality of the data — specifically, their reliability \cite{krippendorff2004content}. Are the annotations collected reproducible, or are they the result of human idiosyncrasies?

\begin{figure}
\centering
  \includegraphics[width=0.48\textwidth]{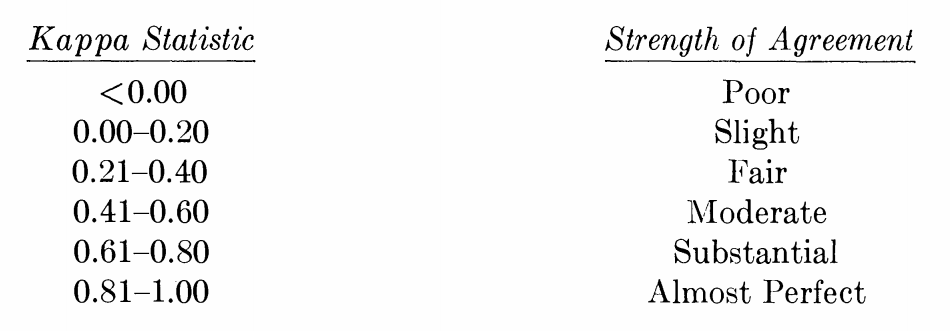}
  \caption[width=\textwidth]{Agreement measures for categorical data \cite{Landis:1977}}
  \label{fig:range}
\end{figure}

Respectable scholarly journals typically require reporting quantitative evidence for the inter-rater reliability (IRR) of the data \cite{Hallgren:2012}. Cohen’s kappa \cite{cohen1960kappa} or Krippendorff’s alpha \cite{hayes2007alpha} is expected to be above a certain threshold to be worthy of publication, typically 0.6 \cite{Landis:1977}. Similar IRR requirements for human annotations data have been followed across many fields. In this paper we refer to this absolute interpretation of IRR as the Landis-Koch approach (Fig. \ref{fig:range}).

This approach has been foundational in guiding the development of widely used and shared datasets and resources. Meanwhile, the landscape of human annotations collection has witnessed a tectonic shift in recent years. Driven by the data-hungry success of machine learning \cite{lecun2015deep, schaekermann2020bet}, there has been an explosive growth in the use of \textit{crowdsourcing} for building datasets and benchmarks \cite{Snow:2008, kochhar2010anatomy}. We identify \textit{three paradigm shifts} in the scope of and methodologies for data collection that make the Landis-Koch approach not as useful in today’s settings.

\paragraph{A rise in annotator diversity} In the pre-crowdsourcing era lab settings, data were typically annotated by two graduate students following detailed guidelines and working with balanced corpora. Over the past two decades, however, the bulk of data are annotated by crowd workers with high cultural and training variances.

\paragraph{A rise in task diversity} There has been an increasing amount of subjective tasks with genuine ambiguity: judging toxicity of online discussions \cite{Aroyo:2019}, in which the IRR values range between 0.2 and 0.4; judging emotions expressed by faces \cite{Cowen:2017}, in which more than 80\% of the IRR values are below 0.6; and A/B testing of user satisfaction or preference evaluations \cite{kohavi2017ab}, where IRR values are typically between 0.3 and 0.5.

\paragraph{A rise in imbalanced datasets} Datasets are no longer balanced intentionally. Many high-stakes human judgements concern rare events with substantial tail risks: event security, disease diagnostics, financial fraud, etc. In all of these cases, a single rare event can be the source of considerable cost. High class imbalance has led to many complaints of IRR interpretability \cite{Byrt:1993,feinstein1990high, cicchetti1990high}.

Each of these changes individually has a profound impact on data reliability. Together, they have caused a shift from  \textit{data-from-the-lab} to  \textit{data-from-the-wild,} for which the Landis-Koch approach to interpreting IRR is admittedly too rigid and too stringent. Meanwhile, we have seen a  drop in the reliance on reliability. Machine learning, crowdsourcing, and data research papers and tracks have abandoned the use and reporting of IRR for human labeled data,  despite calls for it \cite{Paritosh:2012}. The most cited recent datasets and benchmarks used by the community such as SQuAD \cite{rajpurkar2016squad}, ImageNet \cite{deng2009imagenet}, Freebase \cite{bollacker2008freebase}, have never reported IRR values. This would have been unthinkable twenty years ago. More importantly, this is happening against the backdrop of a reproducibility crisis in artificial intelligence \cite{hutson2018crisis}.

With the decline of the usage of IRR, we have seen a rise of ad hoc, misguided quality metrics that took its place, including 1) agreement-\%, 2) accuracy relative to consensus, 3) accuracy relative to “ground truth.” This is dangerous, as IRR is still our best bet for ensuring data reliability. How can we ensure its continued importance in this new era of data collection?

This paper is an attempt to address this problem by proposing an empirical alternative to interpreting IRR. Instead of relying on an absolute scale, we benchmark an experiment's IRR against two baseline measures, to be found in a replication. Replication here is defined as re-annotating the same set of items with a slight change in the experimental setup, e.g., annotator population, annotation guidelines, etc. By fixing the underlying corpus, we can ensure the baseline measures are \textit{sensitive} to the experiment on hand. The first baseline measure is the annotator reliability in the replication. The second measure is the annotator reliability \textit{between} the replications. In Section \ref{sec:xrr}, we present a novel way of measuring this. We call it cross-kappa ($\kappa_x$). It is an extension of Cohen's (1960) kappa and is designed to measure annotator agreement between two replications in a chance-corrected manner.

We present in Appendix \ref{sec:dataset} the International Replication (IRep) dataset,\footnote{https://github.com/google-research-datasets/replication-dataset} a large-scale crowdsourced dataset of four million judgements of human facial expressions in videos. The dataset consists of three replications in Mexico City, Budapest, and Kuala Lumpur.\footnote{On this task, raters received average hourly wages of \$12, \$20, and \$14 USD in Mexico City, Budapest, and Kuala Lumpur respectively. See Appendix \ref{sec:dataset} for annotation setup.} Our analysis in Section \ref{sec:usecase} shows this empirical approach enables meaningful interpretation of IRR. In Section \ref{sec:implications}, we argue xRR is a sensible way of measuring the goodness of crowdsourced datasets, where high reliability is unattainable. While we only illustrate comparing annotator populations in this paper, the methodology behind the xRR framework is general and can apply to similarly replicated datasets, e.g., via change of annotation guidelines.

\section{Related Work}
\label{sec:rel_work}
To position our research, we present a brief summary of the literature in two areas: metrics for measuring annotator agreement and their shortcomings (Section \ref{sec:rel_irr}), comparing replications of an experiment (Section \ref{sec:aggreement_comparison}).  

\subsection{Annotator Agreement}
\label{sec:rel_irr}
\citet{artstein:2008} present a comprehensive survey of the literature on IRR metrics used in linguistics. \citet{popping1988agreement} compare an astounding 43 measures for nominal data (mostly applicable to reliability of data generated by only two observers). Since then, Cohen’s (1960) kappa and its variants \cite{carletta:1997, cohen:1968} have become the de facto standard for measuring agreement in computational linguistics. 

One of the strongest criticisms of kappa is its lack of interpretability when facing class imbalance. This problem is known as the \textit{kappa paradox} \cite{feinstein1990high, Byrt:1993, warrens:2010}, or the ‘base rates’ problem \cite{uebersax:1987}. \citet{bruckner2006} show class imbalance imposes practical limits on kappa and suggest one to interpret kappa \textit{in relation} to the class imbalance of the underyling data. Others have proposed measures that are more robust against class imbalance \cite{gwet2008computing, spitznagel:1985, stewart:1988}. \citet{pontius2011death} even suggest abandoning the use of kappa altogether. 

\subsection{Agreement Between Replications}
\label{sec:aggreement_comparison}
Replications are often being compared, but it is done at the level of per-item mean scores. \citet{Cowen:2017} measure the correlation between the mean scores of two geographical rater pools. They use Spearman’s (1904) correction for attenuation (discussed later in this paper) with split-half reliability. \citet{Snow:2008} measure the Pearson correlations between the score of a single expert and the mean score of a group of non-experts, and vice versa. In this comparison the authors do not correct for correlation attenuation, hence the reported correlations may be strongly biased. Bias aside, correlation is not suitable for tasks with non-interval data or task with missing data. In this paper, we propose a general methodology for measuring rater agreement between replications with the same kind of generality, flexibility, and ease of use as IRR. 

\section{Cross-replication Reliability (xRR)}
\label{sec:xrr}
Data reliability can be assessed when a set of items are annotated multiple times. When this is done by a single rater, \textit{intra-rater reliability} assesses a person’s agreement with oneself. When this is done by two or more raters, \textit{inter-rater reliability} (IRR) assesses the agreement between raters in an experiment. We propose to extend IRR to measure a similar notion of rater-rater agreement, but where the raters are taken from two different experiments. We call it \textit{cross-replication reliability} (xRR). These replications can be a result of re-labeling the same items with a different rater pool, annotation template, or on a different platform, etc.

We begin with a general definition of Cohen's (1960) kappa. We extend it to \textit{cross-kappa} ($\kappa_x$) to measure cross-replication reliability. We then use this foundation to define \textit{normalized} $\kappa_x$ to measure similarity between two replications.

\subsection{Kappa and Its Generalizations}
\label{sec:generalized_kappa}
The class of IRR measures is quite diverse, covering many different experimental scenarios, e.g., different numbers of raters, rating scales, agreement definitions, assumptions about rater interchangeability, etc. Out of all such coefficients, Cohen’s (1960) kappa has a distinct property that makes it most suitable for the task on hand. Unlike Scott’s pi \cite{scott:1955}, Fleiss’s kappa \cite{fleiss:1971}, Krippendorf’s alpha \cite{krippendorff2004content}, and many others, Cohen’s (1960) kappa allows for \textit{two different marginal distributions}. This stems from Cohen’s belief that two raters do not necessarily share the same marginal distribution, hence they should not be treated \textit{interchangeably}. When we compare replications, e.g., two rater populations, we are deliberately changing some underlying conditions of the experiment, hence it is safer to assume the marginal distributions will not be the same. Within either replication, however, we rely on the rater interchangeability assumption. We think this more accurately reflects the current practice in crowdsourcing, where each rater contributes a limited number of responses in an experiment, and hence raters are operationally interchangeable.

Cohen’s (1960) kappa was invented to compare two raters classifying $n$ items into a fixed number of categories. Since its publication, it has been generalized to accommodate multiple raters \cite{light1971, berry:1988}, and to cover different types of annotation scales: ordinal \cite{cohen:1968}, interval \cite{berry:1988, janson2001}, multivariate \cite{berry:1988}, and any arbitrary distance function \cite{artstein:2008}. In this paper we focus on Janson and Olsson’s (2001) generalization, which the authors denote with the lowercase Greek letter iota ($\iota$). It extends kappa to accommodate interval data with multiple raters, and is expressed in terms of pairwise disagreement:
\begin{equation} \label{eq:1.1}
\iota = 1 - \frac{d_o}{d_e}.
\end{equation}
$d_{o}$ in this formula represents the observed portion of disagreement and is defined as:
\begin{equation} \label{eq:1.2}
d_{o} = \Bigg[ n \binom{b}{2} \Bigg]^{-1}\sum_{r<s}\sum_{i}^{n}D(x_{ri},x_{si}),
\end{equation}
where $n$ is the number of items, $b$ the number of annotators, $i$ the item index, $r$ and $s$ the annotator indexes; $\sum_{r<s}$ is the sum over all $r$ and $s$ such that $1<=r<s<=b$. $D()$ is a pairwise disagreement defined as:
\begin{equation} \label{eq:1.3}
D(x_{ri},x_{si})=(x_{ri}-x_{si})^{2}
\end{equation}
for interval data, and
\begin{equation} \label{eq:1.4}
D(x_{ri}, x_{si}) = \begin{cases} 0, & x_{ri} = x_{si} \\
1, & \textrm{otherwise} \end{cases}
\end{equation}
for categorical data. Note we are dropping Janson and Olsson’s multivariate reference in $D()$ and focusing on the univariate case. $d_{e}$ in the denominator represents the expected portion of disagreement and is defined as:
\begin{equation} \label{eq:1.5}
d_e = \Bigg[ n^2 \binom{b}{2} \Bigg]^{-1} \sum_{r<s} \sum_i^n \sum_j^n D(x_{ri}, x_{sj}).
\end{equation}

Janson and Olsson's expression in Eq. \ref{eq:1.1} is based on \citet{berry:1988}. While the latter use absolute distance for interval data, the former use squared distance instead. We follow Janson and Olsson's approach because squared distance leads to desirable properties and familiar interpretation of coefficients \cite{fleiss:1973,krippendorff1970bivariate}. Squared distance is also used in alpha \cite{krippendorff2004content}. \citet{berry:1988} show if $b=2$ and the scale is categorical, $\iota$ in Eq. \ref{eq:1.1} reduces to Cohen’s (1960) kappa. For other rating scales such as ratio, rank, readers should refer to \citet{krippendorff2004content} for additional distance functions. The equations for $d_{o}$ and $d_{e}$ are unaffected by the choice of $D()$.

\subsection{Definition of $\kappa_x$}
\label{sec:cross_kappa}
Here we present $\kappa_x$ as a novel reliability coefficient for measuring the rater agreement between two replications. In Janson and Olsson's generalized kappa above, the disagreement is measured within pairs of annotations taken from the same experiment. In order to extend it to measure cross-replication agreement, we construct annotation pairs such that the two annotations are taken from different replications. We do not consider annotation pairs from the same replication. We define \textit{cross-kappa}, $\kappa_{x}(X,Y)$, as a reliability coefficient between replications $X$ and $Y$:
\begin{equation} \label{eq:2.1}
\kappa_x(X,Y) = 1 - \frac{d_o(X,Y)}{d_e(X,Y)},
\end{equation}
where
\begin{equation} \label{eq:2.2}
d_o(X,Y) = \frac{1}{n R S} \sum_{i=1}^n \sum_{r=1}^R \sum_{s=1}^S D(x_{ri}, y_{si}),
\end{equation}
and
\begin{equation} \label{eq:2.3}
d_e(X,Y) = \frac{1}{n^2 R S} \sum_{i=1}^n \sum_{j=1}^n \sum_{r=1}^R \sum_{s=1}^S D(x_{ri}, y_{sj}),
\end{equation}
where $x$ and $y$ denote annotations from replications $X$ and $Y$ respectively, $n$ is the number of items, $R$ and $S$ the numbers of annotations per item in replications $X$ and $Y$ respectively. In this definition, the observed disagreement is obtained by averaging disagreement observed in $n R S$ pairs of annotations, where each pair contains two annotations on the same item taken from two different replications. Expected disagreement is obtained by averaging over all possible $n^2 R S$ cross-replication annotation pairs. When each replication has only 1 annotation per item, and the data is categorical, it is easy to show $\kappa_x$ reduces to Cohen’s (1960) kappa.

$\kappa_x$ is a kappa-like measure, and will have properties similar to kappa's. $\kappa_x$ is bounded between 0 and 1 in theory, though in practice it may be slightly negative for small sample sizes. $\kappa_x = 0$ means there is no discernible agreement between raters from two replications, beyond what would be expected by chance. $\kappa_x = 1$ means all raters between two replications are in perfect agreement with each other, which also implies perfect agreement within either replication.

\subsection{$\kappa_x$ with Missing Data}
\label{sec:missing_data}
As presented, the two replications can have different numbers of annotations per item. However, within either replication, the number of annotations per item is assumed to be fixed. We recognize this may not always be the case. In practice, items within an experiment can receive varying numbers of annotations (i.e., missing data). We now show how to calculate $\kappa_x$ with missing data.

When computing IRR with missing data, weights can be used to account for varying numbers of annotations within each item. \citet{janson:2004} propose a weighting scheme for iota in Eq. \ref{eq:1.1}. Instead, we follow the tradition of \citet{krippendorff2004content} in weighting each annotation equally in computing $d_{o}$ and $d_{e}$. That amounts to the following scheme. In $d_{o}$, we first normalize within each item, then we take a weighted average over all items, with weights proportional to the combined numbers of annotations per item. In $d_{e}$, no weighting is required.

Since $R$ and $S$ can now vary from item to item, we index them using $R(*)$ and $S(*)$ to denote that they are functions of the underlying items. We rewrite $d_{o}$ and $d_{e}$ as:
\begin{equation} \label{eq:3.1}
d_o(X,Y) = \sum_{i=1}^n \frac{R(i) + S(i)}{\mathbf{R} + \mathbf{S}} \sum_{r=1}^{R(i)} \sum_{s=1}^{S(i)} \frac{D(x_{ri}, y_{si})}{R(i) \cdot S(i)}
\end{equation}
and
\begin{equation} \label{eq:3.3}
d_e(X,Y) = \frac{1}{\mathbf{R} \cdot \mathbf{S}} \sum_{i=1}^n \sum_{j=1}^n \sum_{r=1}^{R(i)} \sum_{s=1}^{S(j)} D(x_{ri}, y_{sj}),
\end{equation}
with
\begin{equation} \label{eq:3.2}
\mathbf{R}=\sum_i^n R(i), \quad \mathbf{S}=\sum_j^n S(j),
\end{equation}
where $\mathbf{R}$ is the total number of annotations in replications $X$, $R(i)$ the number annotations on item $i$ in replication $X$, $r=1, 2, \ldots, R(i)$ (on item $i$ in replication $X$); and similarly for $\mathbf{S}$, $S(j)$, and $s$ with respect to replication $Y$. $\sum_{r=1}^{R(i)}$ and $\sum_{s=1}^{S(j)}$ in Eq. \ref{eq:3.1} and \ref{eq:3.3} are inner summations, where $i$ and $j$ are indexes from the outer summations. Without missing data, $R(i)=R$ for all $i$, and $S(j)=S$ for all $j$, then $\mathbf{R}=nR$, $\mathbf{S}=nS$, reducing Eq. \ref{eq:3.1} and \ref{eq:3.3} to Eq. \ref{eq:2.2} and \ref{eq:2.3}.

\subsection{Normalization of $\kappa_x$}
\label{sec:normalization}
xRR is modeled closely after IRR in order to serve as its baseline. As IRR measures the agreement between raters, so does xRR. In other words, $\kappa_x$ is really a measure of \textit{rater agreement}, not a measure of \textit{experimental similarity} per se. This distinction is important. If we want to measure how well we replicate an experiment, we need to measure its disagreement with the replication \textit{in relationship} to their own internal disagreements. The departure between inter-experiment and intra-experiment disagreements is important in measuring experimental similarity.

This calls for a normalization that considers $\kappa_x$ in relation to IRR. First, we take inspirations from Spearman’s correction for attenuation \cite{spearman:1904}:

\begin{equation} \label{eq:4.1}
\rho_{xy}=\frac{r_{xy}}{\sqrt{reliability_x}\sqrt{reliability_y}},
\end{equation}
where $r_{xy}$ is the observed Pearson product-moment correlation between $x$ and $y$ (variables observed with measurement errors), $\rho_{xy}$ is an estimate of their true, unobserved correlation (in the absence of measurement errors), and $reliability_x$ and $reliability_y$ are the reliabilities of $x$ and $y$ respectively. Eq. \ref{eq:4.1} is Spearman's attempt to correct for the negative bias in $r_{xy}$ caused by the observation errors in $x$ and $y$.\footnote{Spearman relied on the assumption that the errors are uncorrelated with each other and with $x$ and $y$.}

Eq. \ref{eq:4.1} is relevant here because of the close connection between Cohen's (1960) kappa and the Pearson correlation, $r_{xy}$. In the dichotomous case, if the two marginal distributions are the same, Cohen's (1960) kappa is equivalent to the Pearson correlation \cite{cohen1960kappa,cohen:1968}. In the multi-category case, \citet{cohen:1968} generalizes this equivalence to weighted kappa, under the conditions of equal marginals and a specific quadratic weighting scheme.

Based on this strong connection, we propose replacing $r_{xy}$ in Eq. \ref{eq:4.1} with $\kappa_x$ and define normalized $\kappa_x$ as:
\begin{equation} \label{eq:4.2}
\textrm{normalized} \,\kappa_x=\frac{\kappa_x(X,Y)}{\sqrt{\textrm{IRR}_X} \sqrt{\textrm{IRR}_Y}}.
\end{equation}
Defined this way, one would expect normalized $\kappa_x$ to behave like $\rho_{xy}$. That is indeed the case. When we apply both measures to the IRep dataset, we obtain a Pearson correlation of 0.99 between them (see Section \ref{sec:connection_to_rho}). This leads to two insights. First, we can interpret normalized $\kappa_x$ like a disattenuated correlation, $\rho_{xy}$ (see \cite{muchinsky:1996} for a rigorous interpretation). Second, normalized $\kappa_x$ approximates the true correlation between two experiments' item-level mean scores.

Despite their affinity, $\rho_{xy}$ is not a substitute for normalized $\kappa_x$ for measuring experimental similarity. Normalized $\kappa_x$ is more general as it can accommodate non-interval scales and missing data.

\subsection{Connection between xRR and IRR}
\label{sec:connection}
By connecting normalized $\kappa_x$ to $\rho_{xy}$, we can also learn a lot about $\kappa_x$ itself. To the extent that normalized $\kappa_x$ approximates $\rho_{xy}$, we can rewrite Eq. \ref{eq:4.2} as:
\begin{equation} \label{eq:4.3}
\kappa_x(X,Y) \approx \rho_{xy} \sqrt{\textrm{IRR}_X} \sqrt{\textrm{IRR}_Y}.
\end{equation}
This formulation shows $\kappa_x$ behaves like a product of $\rho_{xy}$ and the geometric mean of the two IRRs. This has important consequences, as we can deduce the following. 1) Holding constant the mean scores, and hence $\rho_{xy}$, the lower the IRRs, the lower the $\kappa_x$. Intra-experiment disagreement inflates inter-experiment disagreement.  2) In theory $\rho_{xy}\leq 1.0$,\footnote{Spearman's correction can occasionally produce a correlation above 1.0 \cite{muchinsky:1996}.} hence $\kappa_x$ is capped by the greater of the two IRRs. I.e., Intra-experiment agreement presents a ceiling to inter-experiment agreement. 3) If $x$ and $y$ are identically distributed, e.g., in a perfect replication, $\rho_{xy}=1$ and $\kappa_x(X,Y)=\textrm{IRR}_X=\textrm{IRR}_Y$. Thus, when a low reliability experiment is replicated perfectly, $\kappa_x$ will be as low,
whereas normalized $\kappa_x$ will be 1. This explains why normalized $\kappa_x$ is more suitable for measuring experimental similarity.

In this section, we propose $\kappa_x$ as a measure of rater agreement between two replications, and normalized $\kappa_x$ is as an experimental similarity metric. In the next section, we apply them in conjunction with IRR to illustrate how we can gain deeper insights into experiment reliabilities by triangulating these measures.

\section{Applying xRR to the IRep Dataset}
\label{sec:usecase}
As a standalone measure, IRR captures the reliability of an experiment by encapsulating many of its facets: class imbalance, item difficulty, guideline clarity, rater qualification, task ambiguity, etc. As such, it is difficult to compare the IRR of different experiments, or to interpret their individual values, because IRR is tangled with all the aforementioned design parameters. For example, we cannot attribute a low IRR to rater qualification without first isolating other design parameters. This is the problem we try to solve with xRR by contextualizing IRR with meaningful baselines via a replication. We will demonstrate this by applying this technique to the IRep Dataset (Appendix \ref{sec:dataset}). We focus on a subset of 5 emotions for illustration purposes, with the rest of the reliability values provided in Appendix \ref{sec:full_table}. In our analysis, IRR is measured with Cohen's (1960) kappa and xRR with $\kappa_x$. We will refer to them interchangeably.

\subsection{IRR Variability Across Emotions}
\label{sec:irr_variability_emotions}
\begin{figure}[hbt!]
\centering
  \includegraphics[width=0.48\textwidth]{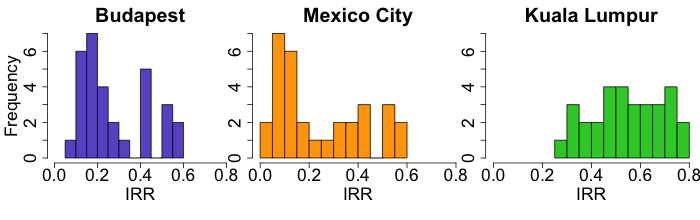}
  \caption{Histograms of 31 emotion labels' IRR in 3 cities. The x-axis denotes buckets of IRR values. The y-axis denotes the number of emotion labels in each of those buckets. There is a lot of variation between emotion labels within each city.}
  \label{fig:hist_irr}
\end{figure}

\begin{figure}[hbt!]
\centering
  \includegraphics[width=0.48\textwidth]{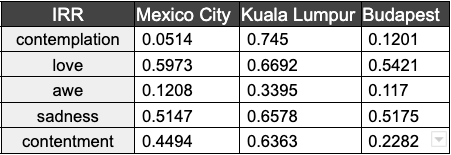}
  \captionof{table}{IRR values of 5 emotion labels in 3 cities.}
  \label{tab:table_irr}
\end{figure}

First we illustrate in Fig. \ref{fig:hist_irr} that different emotions within the same city can have very different IRR. For instance, the labels \textit{awe} and \textit{love} in  Mexico City have an IRR of 0.1208 and 0.597 respectively (Table \ref{tab:table_irr}). \textit{Awe} and \textit{love} are completely different emotions with different levels of class imbalance and ambiguity, and without controlling for these differences, the gap in their reliabilities is not unexpected. That is exactly the problem about comparing IRRs -- such comparisons are not meaningful. We need something directly comparable to \textit{awe} in order to interpret its low IRR. If we do not compare emotions, and just consider \textit{awe} using the Landis-Koch scale, that would not be helpful either. We would not be able to tell if its low IRR is a result of poor guidelines, general ambiguity in emotion detection, or ambiguity specific to \textit{awe}. It's more meaningful to compare replications of \textit{awe} itself.

\subsection{IRR Variability Across Replications}
\label{sec:irr_variability_rep}
\begin{figure}[hbt!]
\centering
  \includegraphics[width=0.48\textwidth]{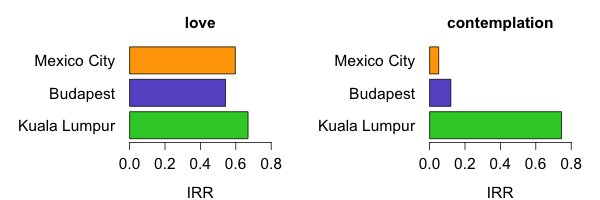}
  \caption{IRR values for label \textit{love} (left) and \textit{contemplation} (right) across the 3 cities. There are different degrees of IRR variability in the two emotion labels.}
  \label{fig:love+contemplation}
\end{figure}

While the aforementioned variation in IRR between emotions is expected, IRR of the \textit{same emotion} can vary greatly between replications as well. Fig. \ref{fig:love+contemplation} shows two contrasting examples. On the one hand, the IRR of \textit{love} is consistent across replications. On the other hand, the IRR of \textit{contemplation} varies a lot. We know the IRR variation in \textit{contemplation} is strictly attributed to rater pool differences because the samples, platforms and annotation templates are the same across experiments. Such variation in IRR will be missed entirely by sampling based approaches for error-bars (e.g. standard error, bootstrap), which assume a fixed rater population.

\subsection{Cross-replication Rater Agreement}
\label{sec:xrr_for_irr}
As shown, replication can facilitate comparisons of IRR by producing meaningful baselines. However, IRR is an internal property of a dataset, it does not allow us to compare two datasets directly. To that end, we can apply $\kappa_x$ to quantify the rater agreement between two datasets, as IRR quantifies the rater agreement within a dataset. Interestingly, not only is $\kappa_x$ useful for comparing two datasets, but it also serves as another baseline for interpreting their IRRs.

IRR is a step toward ensuring reproducibility, so naturally we wonder how much of the observed IRR is tied to the specific experiment and how much of it generalizes? This is of particular concern when raters are sampled in a clustered manner, e.g., crowd workers from the same geographical region, grad students sharing the same office. We rarely make sure raters are diverse and representative of the larger population. High IRR can be the result of a homogeneous rater group, limiting the generality of the results. In the context of the IRep dataset, that two cities having similar IRRs does not imply their raters agree with each other at a comparable level, or at all. We will demonstrate this with two contrasting examples.

\begin{figure}[hbt!]
\centering
  \includegraphics[width=0.3\textwidth]{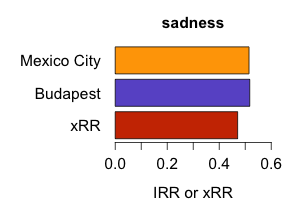}
  \vspace{2mm}
  \caption{IRR values of \textit{sadness} in Mexico City and Budapest and their $\kappa_x$ value. Both cities have as much internal agreement as cross-replication agreement.}
  \label{fig:sadness}
\end{figure}

\begin{figure}[hbt!]
\centering
  \includegraphics[width=0.3\textwidth]{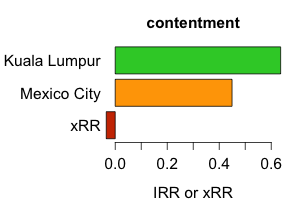}
  \vspace{2mm}
  \caption{IRR of \textit{contentment} in Kuala Lumpur and Mexico City and their $\kappa_x$. Both cities have high internal agreement, but no discernible cross-replication agreement.}
  \label{fig:contentment}
\end{figure}

Mexico City and Budapest both have a moderate IRR for \textit{sadness}, 0.5147 and 0.5175 respectively, and their $\kappa_x$ is nearly the same at 0.4709 (Fig. \ref{fig:sadness}). This gives us confidence that the high IRR of \textit{sadness} generalizes beyond the specific rater pools.
In contrast, on \textit{contentment} Mexico City and Kuala Lumpur have comparable levels of IRR, 0.4494 and 0.6363 respectively, but their $\kappa_x$ is an abysmal -0.0344 \footnote{Negative xRR value due to estimation error.} (Fig. \ref{fig:contentment}). In other words, the rater agreement on \textit{contentment} is limited to within-pool observations only. This serves as an important reminder that IRR is a property of a specific experimental setup and may or may not generalize beyond that. $\kappa_x$ allows us to ensure the internal agreement has external validity.

\subsection{Replication Similarity}
\label{sec:xrr_replications}
\begin{figure}[hbt!]
\centering
  \includegraphics[width=0.3\textwidth]{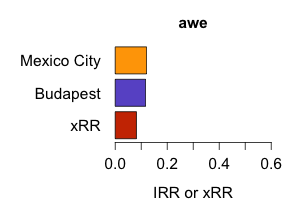}
  \vspace{2mm}
  \caption{IRR of \textit{awe} in Mexico City and Budapest and their xRR. The low xRR is primarily a reflection of their low IRRs.}
  \label{fig:awe}
\end{figure}

\begin{figure}[hbt!]
\centering
  \includegraphics[width=0.48\textwidth]{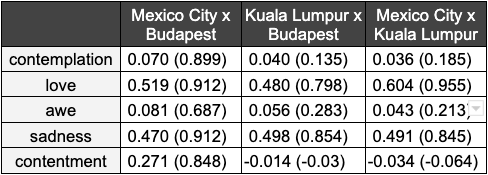}
  \vspace{1mm}
  \captionof{table}{$\kappa_x$ and normalized $\kappa_x$ (in parentheses) of 5 emotion labels in 3 replication pairs.}
  \label{tab:table_xrr}
\end{figure}

$\kappa_x$ is a step towards comparing two replications, but it is not a good standalone measure of replication similarity. To do that, we must also account for both replications' internal agreements, e.g., via normalized $\kappa_x$ in Eq. \ref{eq:4.2}. Fig. \ref{fig:awe} shows an example. Mexico City and Budapest have a low $\kappa_x$ of 0.0817 on \textit{awe}. On the surface, this low agreement may seem attributable to differences between the rater pools. However, there is a similarly low IRR in either city: 0.1208 in Mexico City, and 0.117 in Budapest. After accounting for IRR, normalized $\kappa_x$ is much higher at 0.6872 (Table \ref{tab:table_xrr}), indicating a decent replication similarity between the two cities.

\vspace{2mm}

\subsection{Connection to $\rho_{xy}$}
\label{sec:connection_to_rho}
\begin{figure}[hbt!]
\centering
  \includegraphics[width=0.35\textwidth]{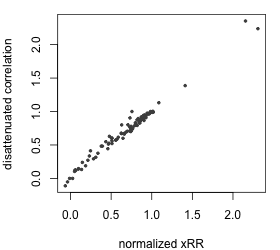}
  \vspace{2mm}
  \caption{Scatter plot of $\rho_{xy}$ (y-axis) and normalized $\kappa_x$ (x-axis). Each dot is an emotion label in a pair of replications.}
  \label{fig:split_half}
\end{figure}

We apply Spearman's correction for attenuation in Eq. \ref{eq:4.1} to all 31 emotion labels in 3 replication pairs. The resulting $\rho_{xy}$ is plotted against the corresponding normalized $\kappa_x$ in Fig. \ref{fig:split_half}. Both measures are strongly correlated with a Pearson correlation of 0.99. This justifies interpreting normalized $\kappa_x$ as a disattenuated correlation like $\rho_{xy}$.

\section{Measuring the Quality of a Crowdsourced Dataset}
\label{sec:implications}
The IRep dataset is replicated and is conducive to xRR analysis. However, in practice most datasets are not replicated. Is xRR still useful? We present a specific use case of xRR in this section and argue that it is worth replicating a crowdsourced dataset in order to evaluate its quality.

\subsection{Data Target}
\label{sec:goodness}
Given a set of items, it is possible that annotations of the highest attainable quality still fail to meet the Landis-Koch requirements. Task subjectivity and class imbalance together impose a practical limit on kappa \cite{bruckner2006}. In these situations, the experimenter can forgo a data collection effort for reliability reasons. Alternatively, the experimenter may believe that data of sufficiently high quality can still have scientific merits, regardless of reliability. If so, what guidance can we use to \textit{ensure the highest quality data}, especially when collecting data via crowdsourcing? This paper is heavily motivated by this question.

xRR allows us to interpret IRR not on an absolute scale, but against a replication, a reference of sorts. By judging a crowdsourced dataset against a reference, we can decide if its meets a certain quality bar, albeit a relative one. In the IRep dataset, all replications are of equal importance. However, in practice, we can often define a \textit{trusted source} as our target. This trusted source can consist of linguists, medical experts, calibrated crowd workers, or the experimenters themselves. They should have enough expertise knowledge and an adequate understanding of the task. The critical criterion in choosing a target is its ability to remove common quality concerns such as rater qualification and guideline effectiveness.

\subsection{Internal Agreements}
By replicating a random subset of a crowdsourced dataset with trusted annotators,\footnote{2 or more ratings per item are needed to measure the IRR.} one can compare the two IRRs and make sure they are at a similar level. If the crowd IRR is much higher, that may be an indication of collusion, or a set of overly simplistic guidelines that have deviated from the experiment fidelity \cite{sameki2015rigorously}. If the crowd IRR is much lower, it may just be a reflection of annotator diversity, or it can mean under-defined guidelines, unequal annotator qualifications, etc. Further investigation is needed to ensure the discrepancy is reasonable and appropriate.

\subsection{Mutual Agreement}
Suppose the two IRRs are similar, that is not to say that both datasets are similar. Both groups of annotators can have high internal agreement amongst themselves, but the two groups can agree on different sets of items. If our goal is to collect crowdsourced data that closely mirror the target, then we have to measure their mutual agreement, in addition to comparing their internal agreements. Recall from Section \ref{sec:connection} that if an experiment is replicated perfectly, $\kappa_x$ should be identical to the two IRRs. Or more concisely, normalized $\kappa_x$ should be equal to 1. Thus a high normalized $\kappa_x$ can assure us that the crowdsourced annotators are functioning as an extension of the trusted annotators, based on which we form our expectations.

\subsection{Relation to Gold Data}
At a glance, this approach seems similar to the common practice of measuring the accuracy of crowdsourced data against the ground truth \cite{resnik:2006,hripcsak:2002}. However, they are actually fundamentally different approaches. $\kappa_x$ is rooted in the reliability literature that does not rely on the existence of a \textit{correct} answer. The authors argue this is an unrealistic assumption for many crowdsourcing tasks, where the input involves some subjective judgement. Accuracy itself is also a flawed metric for annotations data due to its inability to handle data uncertainty. For instance, when the reliability of the gold data is less than perfect, accuracy can never reach 1.0. Furthermore, accuracy is not chance-corrected, so it tends to inflate with class imbalance.

\subsection{Extending an Existing Dataset}
\label{sec:extending}
The aforementioned technique can also measure the quality of a dataset extension. The main challenge in extending an existing dataset is to ensure the new data is consistent with the old. The state-of-the-art method in computer vision is \textit{frequency matching} -- ensuring the same proportion of yes/no votes in each image class. \citet{recht:2019} extended ImageNet\footnote{\url{http://www.image-net.org/}} using this technique, concluding there is a 11\% - 14\% drop in accuracy across a broad range of models. While frequency matching controls the distribution of some statistics, the impact of the new platform is uncontrolled for. \citet{engstrom:2020} pointed out a bias in this sampling technique. Overall, it is difficult to assess how well we are extending a dataset. To that end, xRR can be of help. A high normalized $\kappa_x$ and a comparable IRR in the new data can give us confidence in the uniformity and continuity in the data collection.

\section{Discussion}
\label{discussion}
There has been a tectonic shift in the scope of and methodologies for annotations data collection due to the rise of crowdsourcing and machine learning. In many of these tasks, a high reliability is often difficult to attain, even under favorable circumstances. The rigid Landis-Koch scale has resulted in a decrease in the usage and reporting of IRR in most widely used datasets and benchmarks. Instead of abandoning IRR, we should adapt it to new ways of measuring data quality. The xRR framework presents a first-principled way of doing so. It is a more empirical approach that utilizes a replication as a reference point. It is based on two metrics -- $\kappa_x$ for measuring cross-replication rater agreement -- and normalized $\kappa_x$ for measuring replication similarity.

We opensource a large-scale replication dataset of facial expression judgements analyzed with the proposed framework. We show this framework can be used to guide our crowdsourcing data collection efforts. This is the beginning of a long line of inquiry. We outline future work and limitations below:

\paragraph{Confidence intervals for $\kappa_x$} Confidence intervals for $\kappa_x$ and normalized $\kappa_x$ are required for hypothesis testing. Though one can use the block-bootstrap for an empirical estimate, large sample behavior of these metrics needs to be studied.

\paragraph{Sensitivity of $\kappa_x$ with high class-imbalance} The xRR framework sidesteps the effect of class-imbalance by comparing replications on the same item set. Further analysis needs to confirm the sensitivity of $\kappa_x$ metrics in high class-imbalance. 

\paragraph{Optimization of $\kappa_x$ computation}  Our method requires constructing many pairs of observations: $n^2 R S$. This may get prohibitively expensive, when the number of items is large. Using algebraic simplification and dynamic programming, this can be made much more efficient.

\paragraph{Alternative normalizations of $\kappa_x$} We provided one particular normalization technique, but it may not suit all applications. For example, when comparing crowd annotations to expert annotations, one can consider, $\kappa_x/$IRR$_{expert}$.

\paragraph{Alternative xRR coefficients} Our proposed xRR coefficient, $\kappa_x$, is based on Cohen's (1960) kappa for its assumption about rater non-interchangeability. It may be useful to consider Krippendorff’s alpha and other agreement statistics as alternatives for other assumptions and statistical characteristics.

\paragraph{} We hope this paper and dataset will spark research on these questions and increase reporting of reliability measures for human annotated data. 

\section*{Acknowledgments}
We like to thank Gautam Prasad and Alan Cowen for their work on collecting and sharing the IRep dataset and opensourcing it.

\bibliographystyle{acl_natbib}
\bibliography{anthology,acl2021}

\appendix
\appendixpage
\section{The IRep Dataset: Facial Expressions Replication Dataset}
\label{sec:dataset}

\begin{figure*}[hb!]
\centering
  \includegraphics[width=0.7\textwidth]{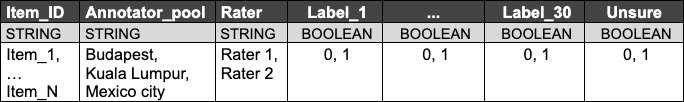}
  \vspace{3mm}
  \captionof{table}{Schema of the CSV file: Each line in the IRep csv file corresponds to one item\_ID annotated by Rater\_1 and Rater\_2 with some of the emotion labels (Label\_1 … Label\_30) annotated on the corresponding. There is one additional column for “unsure” indicating when it was not possible to determine which expression was expressed.}
  \label{fig:irep_schema}
\end{figure*}

The IRep Dataset is a human annotated dataset with a list of 30 emotion labels from a set of emotion classes defined in \citet{Cowen:2017} plus one additional label ‘unsure’. During the data collection process, raters used the set of 30 facial expression labels to annotate their perception of the facial expression present in a video. As the purpose of this dataset is to illustrate how replications of human labeling experiments can be used to determine the overall quality of the resulting annotations, we have omitted the reference to the actual video content. The annotated videos are thus referred to as ‘items’ with a set of indices (item\_IDs), e.g. item\_1, item\_2, etc, stored in a CSV format. Raters are referred to as Rater\_1 or Rater\_2 across all rater pools. They are just placeholders and do not imply particular individuals (Table \ref{fig:irep_schema}).

\begin{figure*}[hb!]
\centering
  \includegraphics[width=0.65\textwidth]{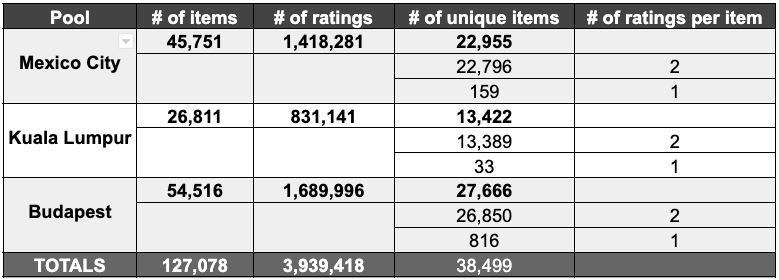}
  \vspace{3mm}
  \captionof{table}{Distribution of items and ratings across different rater pools, where every item is annotated by a maximum of 2 raters from each pool. Here we show what fraction of the unique items were annotated by one or two raters in each rating pool.}
  \label{fig:irep_stats}
\end{figure*}

The dataset contains 3,939,418 annotations on 38,499 unique items. The size of the dataset is 15MB, and the dataset is released on GitHub: \url{https://github.com/google-research-datasets/replication-dataset}. To produce the replications for this labeling experiment, we used rater pools in three different cities - Budapest, Kuala Lumpur and Mexico City - on the same labeling platform. In Table \ref{fig:irep_stats} we show the distribution of items and ratings across the different rater pools.

\section{IRR, xRR, and normalized xRR values for the IRep dataset}
\label{sec:full_table}
\begin{figure*}[hb!]
\centering
  \includegraphics[width=\textwidth]{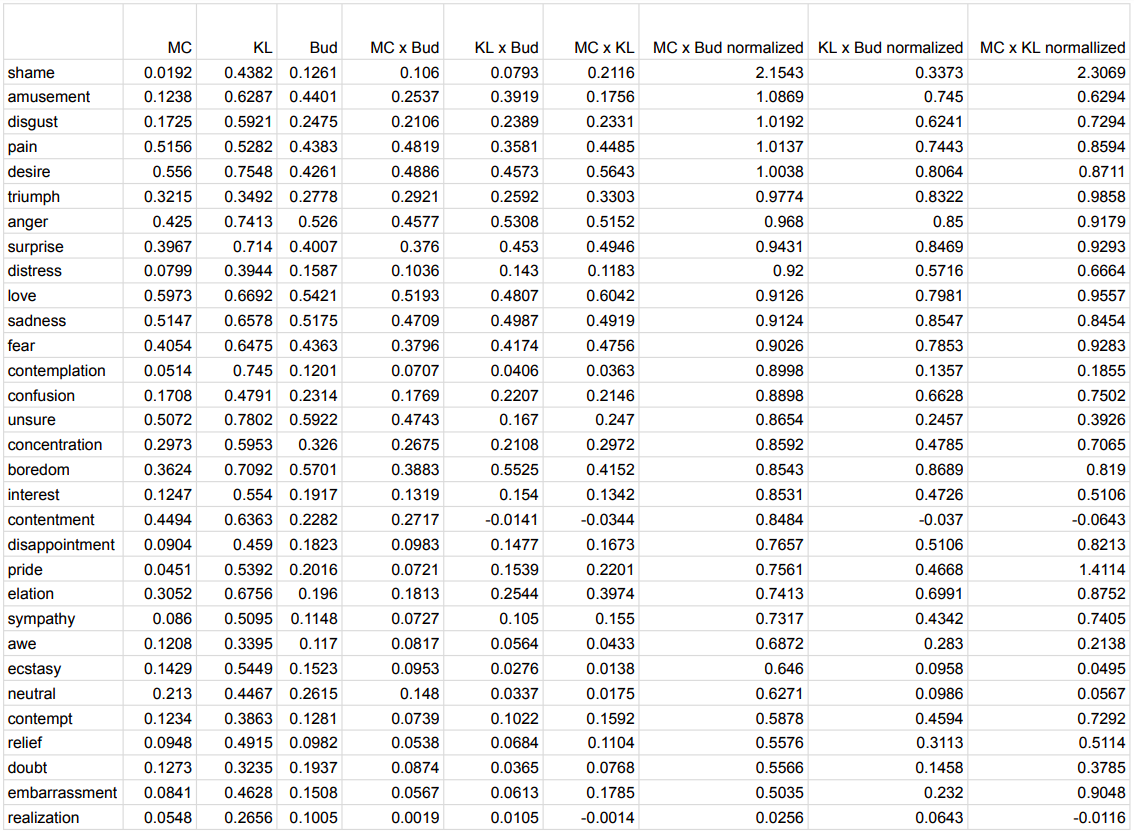}
  \vspace{1mm}
  \captionof{table}{The fist column shows the 30 emotion labels + ``unsure'' in the IRep dataset. The next 3 columns are their IRR measured by Cohen's (1960) kappa in Mexico City (MC), Kuala Lumpur (KL), and Budapest (Bud). The next 3 columns are the $\kappa_x$ in the 3 pairs of cities, and the last 3 columns are the corresponding normalized $\kappa_x$.}
  \label{tab:full_table}
\end{figure*}

In Table \ref{tab:full_table} we report the IRR, $\kappa_x$, and normalized $\kappa_x$ obtained from the entire IRep dataset.

\end{document}